\documentclass[aps,showpacs,preprintnumbers,amsmath,amssymb,nofootinbib]{revtex4}
\usepackage{graphicx}

\begin{document}

\preprint{hep-th/0208033}

\title{Rotational Perturbations of High Density Matter in the Brane Cosmology}

\author{Chiang-Mei Chen} \email{cmchen@fermi.phy.ncu.edu.tw}
\affiliation{Department of Physics, National Taiwan University,
Taipei 106, Taiwan}

\author{T. Harko}
\email{harko@hkucc.hku.hk} \affiliation{ Department of Physics,
The University of Hong Kong, Pokfulam Road, Hong Kong}

\author{W. F. Kao} \email{wfgore@cc.nctu.edu.tw}
\affiliation{Institute of Physics, Chiao Tung University, Hsinchu,
Taiwan}

\author{M. K. Mak} \email{mkmak@vtc.edu.hk}
\affiliation{Department of Physics, The University of Hong Kong,
Pokfulam Road, Hong Kong}

\date{October 14, 2003}

\begin{abstract}
We consider the evolution of small rotational perturbations, with
azimuthal symmetry, of the brane-world cosmological models. The
equations describing the temporal, radial, and angular dependence
of the perturbations are derived by taking into account the
effects of both scalar and tensor parts of the dark energy term
on the brane. The time decay of the initial rotation is
investigated for several types of equation of state of the
ultra-high density cosmological matter. For an expanding
Universe, rotation always decays in the case of the perfect
dragging, for which the angular velocity of the matter on the
brane equals the rotation metric tensor. For non-perfect
dragging, the behavior of the rotation is strongly equation of
state dependent. For some classes of dense matter, like the stiff
causal or the Chaplygin gas, the angular velocity of the matter
on the brane is an increasing function of time. For other types
of the ultra-dense matter, like the Hagedorn fluid, rotation is
smoothed out by the expansion of the Universe. Therefore the
study of dynamics of rotational perturbations of brane world
models, as well as in general relativity, could provide some
insights on the physical properties and equation of state of the
cosmological fluid filling the very early Universe.
\end{abstract}
\maketitle
\section{Introduction}
Most of the astronomical objects in the Universe (planets, stars
or galaxies) have some form of rotation (differential or uniform).
Hence the possibility that the Universe itself could be rotating
has attracted a lot of attention. But even that observational
evidences of cosmological rotation have been reported
\cite{Bi82,Bi83,NoRa97,Ku97}, they are still subject of
controversy.

From the analysis of microwave background anisotropy Collins and
Hawking \cite{CoHa73}, and Barrow, Juszkiewicz, and Sonoda
\cite{BaJuSo85} have found some very tight limits of the
cosmological vorticity, $T_{obs} > 3 \times 10^5 \, T_H$, where
$T_{obs}$ is the actual rotation period of our Universe and $T_H
= (1\sim 2)\times 10^{10}$ years is the Hubble time, corresponding
to an angular velocity of the order of $10^{-13}$ rad/sec.
Therefore our present day Universe is rotating very slowly, if at
all. However, the existence of such a small rotation, when
extrapolated to the early stages of the Universe, could have
played a major role in the dynamics of the early Universe, and
possibly as well in the processes involving galaxy formation.

From a theoretical point of view, G\"{o}del \cite{Go49} gave in
1949 his famous example of a rotating cosmological solution to the
Einstein gravitational field equations. G\"{o}del also discussed
the possibility of a cosmic explanation of the galactic rotation
\cite{Go49}. This rotating solution has attracted considerable
interest because the corresponding Universes possess the property
of closed time-like curves.

The investigation of rotating and rotating-expanding Universes
generated a large amount of literature in the field of general
relativity, the combination of rotation with expansion in
realistic cosmological models being one of the most difficult
tasks in cosmology (see \cite{Ob00} for a review of the
expansion-rotation problem in general relativity). Hence rotating
solutions of the gravitational field equations cannot be excluded
{\em a priori}. But this raises the question of why the Universe
rotates so slowly. This problem can also be naturally solved in
the framework of the inflationary model. Ellis and Olive
\cite{ElOl83}, and Gr{\o}n and Soleng \cite{GrSo87} pointed out
that if the Universe came into being as a mini-universe of Planck
dimensions and went directly into an inflationary epoch driven by
a scalar field with a flat potential, due to the non-rotation of
the false vacuum and the exponential expansion during inflation
the cosmic vorticity has decayed by a factor of about $10^{-145}$.

The rotational perturbations of a spatially homogeneous and
isotropic Universe in terms of a small variation of curvature have
been investigated by Hawking \cite{Ha66}. He found that for
pressureless dust and radiation the perturbations die away. Later
Miyazaki \cite{Mi79} studied the perturbations by a spherical
shell in a closed homogeneous Universe in the framework of the
Brans-Dicke theory. The possibility of incorporating a slowly
rotating Universe into the framework of Friedmann-Robertson-Walker
(FRW) type metrics has been considered by Bayin and Cooperstock
\cite{BaCo80}, who obtained the restrictions imposed by the field
equations on the matter angular velocity. They also shown that
uniform rotation is incompatible with the dust filled (zero
pressure) and with the radiation dominated Universe. Moreover,
Bayin \cite{Ba85} has obtained the solutions of the field
equations for a special class of non-separable rotation functions
of the matter distribution. The investigation of the first order
rotational perturbations of flat FRW type Universes proved to be
useful in the study of string cosmological models with dilaton
and axion fields \cite{ChHaMa01}. The form of the rotation
equation imposes strong constraints on the form of the dilaton
field potential $U$, restricting the allowed forms to two: the
trivial case $U=0$ and the exponential type potential.

Recently, Randall and Sundrum \cite{RS99a,RS99b} have pointed out
that a scenario with an infinite fifth dimension in the presence
of a brane can generate a theory of gravity, which mimics purely
four-dimensional gravity, both with respect to the classical
gravitational potential and with respect to gravitational
radiation. The gravitational self-couplings are not significantly
modified in this model. This result has been obtained from the
study of a single 3-brane embedded in five dimensions, with the
5D metric given by $ds^2=e^{-f(y)} \eta_{\mu\nu} dx^\mu dx^\nu +
dy^2$, which can produce a large hierarchy between the scale of
particle physics and gravity, due to the appearance of the warp
factor. Even if the fifth dimension is uncompactified, standard
4D gravity is reproduced on the brane. This model allows the
presence of large or even infinite non-compact extra dimensions.
Our brane is identified to a domain wall in a 5-dimensional
anti-de Sitter space-time.

The Randall-Sundrum (RS) model was inspired by superstring theory.
The ten-dimensional $E_8\times E_8$ heterotic string theory, which
contains the standard model of elementary particle, could be a
promising candidate for the description of the real Universe. This
theory is connected with an eleven-dimensional theory compactified
on the orbifold $R^{10}\times S^1/Z_2$ \cite{HW96}. The static RS
solution has been extended to time-dependent solutions and their
cosmological properties have been extensively studied
\cite{KK00,BDEL00,BDL00,STW00,LMW00,CEHS00,ClGrSe99,AnNuOl00} (for
a review of dynamics and geometry of brane universes see
\cite{Ma01}).

The effective gravitational field equations on the brane world, in
which all the matter forces except gravity are confined on the
3-brane, in a 5-dimensional space-time with $Z_2$-symmetry have
been obtained, by using an elegant geometric approach, by
Shiromizu, Maeda and Sasaki \cite{SMS00,SSM00}. The correct
signature for gravity is provided by the brane with positive
tension. If the bulk space-time is exactly anti-de Sitter,
generically the matter on the brane is required to be spatially
homogeneous. The electric part of the 5-dimensional Weyl tensor
$E_{IJ}$ gives the leading order corrections to the conventional
Einstein equations on the brane. The effect of the dilaton field
in the bulk can also be taken into account in this approach
\cite{MW00}.

In brane-world model, the behavior of an anisotropic Bianchi type
I cosmology in the presence of inflationary scalar fields has
been considered by Maartens, Sahni and Saini \cite{MSS00}. By
using dynamical systems techniques, the behavior of the FRW,
Bianchi type I and V cosmological models in the RS brane world
scenario, with matter on the brane obeying a barotropic equation
of state, has been studied by Campos and Sopuerta
\cite{CS01,CS01a}. The general exact solution of the field
equations for an anisotropic brane with Bianchi type I and V
geometry, with perfect fluid and scalar fields as matter sources,
has been found in \cite{ChHaMa01a}. In spatially homogeneous
brane world cosmological models the initial singularity is
isotropic, and hence the initial conditions problem is solved
\cite{Co01a}. Consequently, these models do not exhibit Mixmaster
or chaotic-like behavior close to the initial singularity
\cite{Co01b}.

Realistic brane-world cosmological models require the
consideration of more general matter sources to describe the
evolution and dynamics of the very early Universe. The effects of
the bulk viscosity of the matter on the brane have been analyzed
in \cite{ChHaMa01b}. Limits on the initial anisotropy induced by
the 5-dimensional Kaluza-Klein graviton stresses by using the CMB
anisotropies have been obtained by Barrow and Maartens
\cite{BaMa01}. Anisotropic Bianchi type I brane-worlds with a pure
magnetic field and a perfect fluid have also been analyzed
\cite{BaHe01}. The effect of the bulk viscosity of the
cosmological matter on the cosmological evolution on the brane for
a Bianchi type I brane geometry was considered in \cite{HaMa03}.

The simplest way to investigate if brane world cosmologies are
consistent with the observations is to investigate the behavior of
the perturbations in the model. Perturbations on the brane are
associated with perturbations in the geometry of the bulk
space-time. The linearized perturbation equations in the
generalized RS model have been obtained, by using the covariant
nonlinear dynamical equations for the gravitational and matter
fields on the brane, by Maartens \cite{Ma00}. The gauge-invariant
formalism for perturbations in the brane world has been developed
in \cite{KoIsSa00,Ko01}. The equations governing the bulk
perturbations in the case of a general warped Universe have been
computed by Langlois \cite{La00,La01}. A gauge invariant formalism
for metric perturbations in five-dimensional brane world theories,
which also applies to models originating from heterotic M-theory
has been obtained in \cite{BrDoBrLu00}. Koyama and Soda
\cite{KoSo00} obtained a formalism for solving the coupled
dynamics of the cosmological perturbations in the brane world and
of the gravitational waves in the AdS bulk. A closed system of the
perturbation equations on the brane, which is valid on a large
scale and may be solved without solving for the bulk perturbations
has been proposed in \cite{GoMa00} and \cite{LaMaSaWa01}. The
perturbations of the brane worlds in conformally Minkowskian
coordinates, which enable to disentangle the contributions of the
bulk gravitons and of the motion of the brane, have been
considered in \cite{DeDoKa01}. Dorca and van de Bruck
\cite{BrDo01} proposed a new gauge, in which the full
five-dimensional problem of the perturbations is solvable. The
second order perturbations of the gravitational field induced on
the 3-brane have been analyzed by Kudoh and Tanaka
\cite{KuTa01,KuTa01a}. The equations of motion for metric
perturbations in the bulk and matter perturbations on the brane
have been presented, in an arbitrary gauge, in \cite{BrMaWa01}.
The evolution of density perturbations in brane world cosmological
models with a bulk scalar field has been considered by Brax, van
de Bruck and Davis \cite{BrBrDa01}. The $1+3$-covariant approach
to cosmological perturbations in the brane-world models, and its
application to CMB anisotropies, have been reviewed recently by
Maartens \cite{Ma03}.

In a previous paper \cite{ChHaKaMa02}, we studied first order
rotational perturbations of homogeneous and isotropic FRW brane
world cosmological models. Assuming that the rotation is slow, and
by keeping only the first order rotational terms in the field
equations, a rotation equation describing the space dependence and
time evolution of the metric perturbations is obtained. However,
in our previous consideration in \cite{ChHaKaMa02} the bulk
effects related to the tensorial part of the dark energy (coming
from the 5-dimensional Weyl tensor) have been neglected, and only
the role played by the scalar part (via the unperturbed field
equations) has been considered. The conservation of the angular
momentum, following from the general energy-momentum conservation
equation on the brane, has also not been included in the
formalism. Moreover, the angular dependence of the angular
velocity of the matter on the brane has also been overlooked.

It is the purpose of the present paper to consider the rotational
perturbations of slowly rotating brane worlds, with azimuthal
symmetry, by taking into account both the tensorial and scalar
perturbations in the bulk, produced due to the matter
perturbations on the brane. We assume that the background geometry
of the unperturbed system is homogeneous and isotropic, of FRW
type. Since the quadratic corrections terms are related, via a
consistency condition, to the dark energy term, the contributions
of the tensorial and scalar parts of the perturbation can be
consistently included in the formalism. The conservation of the
angular momentum of the matter on the brane gives a basic relation
between the angular velocity $\omega$ of the matter and the metric
rotation function $\Omega$, in terms of the equation of state of
the matter on the brane and the scale factor of the expanding
Universe. In the case of the perfect dragging, when
$\omega=\Omega$, the decay of the rotational perturbations is
independent of the equation of state of the matter and in the
large time limit rotation always vanishes. However, the behavior
of perturbations is very different for the general case
$\omega\neq\Omega$. In such case the analysis of the long time
behavior of the angular velocity for several proposed physical
models of the high-density cosmological fluid, corresponding to
different equations of state of the matter (Zeldovich causal model
or Chaplygin gas) shows that in these cases the rotational
perturbations increase in time. However, for the Hagedorn equation
of state of dense matter, the late time rotational perturbations
tend, in the large time limit, to zero. For other, more
conventional equations of state, like the radiation fluid,
pressureless dust, quark matter obeying the bag model equation of
state and the Boltzmann gas, the rotational perturbations are
wiped out by the expansion of the Universe. It worth to note that
{\em the results for the behavior of perturbations are also
specific to standard general relativity}.

The present paper is organized as follows. The field equations
for a slowly rotating brane-world are written down, and the basic
rotation equation is derived in Section II. The case of the
perfect dragging is discussed in Section III. In Section IV the
behavior of rotational perturbations is analyzed for several
relevant equations of state of the cosmological fluid. We
conclude our results in Section V.

\section{Slowly rotating brane Universes}
In the 5D space-time the brane-world is located at $Y(X^I)=0$,
where $X^I, \, I=0,1,2,3,4$, are 5-dimensional coordinates. The
effective action in five dimensions is \cite{MW00}
\begin{equation}
S = \int d^5X \sqrt{-g_5} \left( \frac1{2k_5^2} R_5 - \Lambda_5
\right) + \int_{Y=0} d^4x \sqrt{-g} \left( \frac1{k_5^2} K^\pm -
\lambda + L^{\text{matter}} \right),
\end{equation}
with $k_5^2=8\pi G_5$ the 5-dimensional gravitational coupling
constant and where $x^\mu, \, \mu=0,1,2,3$, are the induced
4-dimensional brane world coordinates. $R_5$ is the 5D intrinsic
curvature in the bulk and $K^\pm$ is the extrinsic curvature on
either side of the brane.

On the 5-dimensional space-time (the bulk), with the negative
vacuum energy $\Lambda_5$ source of the gravitational field the
Einstein field equations are given by
\begin{equation}
G_{IJ} = k_5^2 T_{IJ}, \qquad T_{IJ} = - \Lambda_5 g_{IJ} +
\delta(Y) \left[ -\lambda g_{IJ} + T_{IJ}^{\text{matter}} \right],
\end{equation}
In this space-time a brane is a fixed point of the $Z_2$
symmetry. In the following capital Latin indices run in the range
$0,...,4$ while Greek indices take the values $0,...,3$.

Assuming a metric of the form $ds^2=(n_I n_J + g_{IJ}) dx^I
dx^J$, with $n_I dx^I=d\chi$ the unit normal to the
$\chi=\text{constant}$ hypersurfaces and $g_{IJ}$ the induced
metric on $\chi=\text{constant}$ hypersurfaces, the effective
four-dimensional gravitational equations on the brane take the
form \cite{SMS00,SSM00}:
\begin{equation}
G_{\mu\nu} = - \Lambda g_{\mu\nu} + k_4^2 T_{\mu\nu} + k_5^4
S_{\mu\nu} - E_{\mu\nu},  \label{EqEinstein}
\end{equation}
where
\begin{equation}
S_{\mu\nu} = \frac1{12}T T_{\mu\nu} - \frac14 T_\mu{}^\alpha
T_{\nu\alpha} + \frac1{24} g_{\mu\nu} \left( 3 T^{\alpha\beta}
T_{\alpha\beta} - T^2 \right),
\end{equation}
and $\Lambda=k_5^2(\Lambda_5+k_5^2 \lambda^2/6)/2,\,
k_4^2=k_5^4\lambda/6$ and $E_{IJ}=C_{IAJB} n^A n^B$. $C_{IAJB}$
is the 5-dimensional Weyl tensor in the bulk and $\lambda$ is the
vacuum energy on the brane. $T_{\mu\nu}$ is the matter
energy-momentum tensor on the brane and $T=T^\mu{}_\mu$ is the
trace of the energy-momentum tensor.

For any matter fields (scalar field, perfect fluids, kinetic
gases, dissipative fluids etc.) the general form of the brane
energy-momentum tensor can be covariantly given as
\begin{equation}
T_{\mu\nu} = (\rho+p) u_\mu u_\nu + p h_{\mu\nu} + \pi_{\mu\nu} +
2 q_{(\mu} u_{\nu)}. \label{EMT}
\end{equation}
The decomposition is irreducible for any chosen 4-velocity
$u^\mu$. Here $\rho$ and $p$ are the energy density and isotropic
pressure, $\pi_{\mu\nu}$ are the anisotropic stresses on the
brane, induced for example by the dissipative properties of the
fluid (shear viscosity) and $q_\mu$ is the heat flux. $h_{\mu\nu}
= g_{\mu\nu} + u_\mu u_\nu$ projects orthogonal to $u^\mu$. The
heat flux obeys the condition $q_\mu=q_{<\mu>}$, while the
anisotropic stress obeys $\pi_{\mu\nu}=\pi_{<\mu\nu>}$, where
angular brackets denote the projected, symmetric and trace-free
part:
\begin{equation}
V_{<\mu>} = h_\mu{}^\nu V_\nu, \qquad W_{<\mu\nu>} = \left[
h_{(\mu}{}^{\alpha} h_{\nu)}{}^\beta - \frac13 h^{\alpha\beta}
h_{\mu\nu} \right] W_{\alpha\beta}.
\end{equation}

The symmetry properties of $E_{\mu\nu}$ imply that generally we
can irreducibly decompose it with respect to a chosen 4-velocity
field $u^\mu$ as
\begin{equation}
E_{\mu\nu} = - \frac{k_5^4}{k_4^4} \left[ \frac13 {\cal U}( 4
u_\mu u_\nu + g_{\mu\nu}) + {\cal P}_{\mu\nu} + 2{\cal Q}_{(\mu}
u_{\nu)} \right],  \label{DefE}
\end{equation}
where ${\cal U}$ is a scalar, ${\cal Q}_\mu$ a spatial vector and
${\cal P}_{\mu\nu}$ a spatial, symmetric and trace-free tensor.
For a FRW model ${\cal Q}_\mu=0$ and ${\cal P}_{\mu\nu}=0$
\cite{CS01a}. Hence the only non-zero contribution from the
5-dimensional Weyl tensor from the bulk is given by the scalar
term ${\cal U}$. The ``dark energy'' term, $E_{\mu\nu}$, is a
pure bulk effect, therefore we cannot determine its expression
without solving the complete system of field equations in 5
dimensions. However, its expression in the bulk is constrained by
the relation
\begin{equation}
D_\mu E^{\mu\nu} = k_5^4 D_\mu S^{\mu\nu}.  \label{EqCon}
\end{equation}
The Einstein equation in the bulk also implies the conservation of
the energy momentum tensor of the matter on the brane,
\begin{equation}
T_\mu{}^\nu{}_{;\nu} \mid _{\chi=0} = 0. \label{EqEMC}
\end{equation}

The rotationally perturbed metric can be expressed in terms of
the usual coordinates in the form \cite{MuFeBr92}
\begin{equation}
ds^2 = - dt^2 + a^2(t) \left[ \frac{dr^2}{1-kr^2} + r^2 \left(
d\theta^2 + \sin^2\theta d\varphi^2 \right) \right] - 2
\Omega(t,r,\theta) a^2(t) r^2 \sin^2\theta \, dt d\varphi,
\label{R6}
\end{equation}
where $\Omega(t,r,\theta)$ is the metric rotation function.
Although $\Omega$ plays a role in the ``dragging'' of local
inertial frames, it is not the angular velocity of these frames,
except for the special case when it coincides with the angular
velocity of the matter fields. $k=1$ corresponds to closed
Universes, with $0\leq r\leq 1$. $k=-1$ corresponds to open
Universes, while the case $k=0$ describes a flat geometry, where
the range of $r$ is $0\leq r<\infty $. In all models, the
time-like variable $t$ ranges from $0$ to $\infty$.

For the matter energy-momentum tensor on the brane we restrict
our analysis to the case of the perfect fluid energy-momentum
tensor,
\begin{equation}
T^{\mu\nu} = (\rho+p) u^\mu u^\nu + p g^{\mu\nu}.
\end{equation}
The components of the four-velocity vector are
$u^\mu=(1,0,0,\omega)$ and $\omega(t,r,\theta)=d\varphi/dt$ is
the angular velocity of the matter distribution. Consequently,
for the rotating brane the energy-momentum tensor has a
supplementary component
\begin{equation}
T_{03} = \left\{ \left[ \Omega(t,r,\theta) - \omega(t,r,\theta)
\right] \rho - \omega(t,r,\theta) p \right\} r^2 a^2(t) \sin^2
\theta.
\end{equation}
As a first step to consider rotational perturbations on the
brane, we should find a consistent assumption for $E_{\mu\nu}$
``induced'' by the small rotation. The reason that the small
rotation changes the expression of $E_{\mu\nu}$ can be understood
as follows. In the brane world scenario an $Z_2$ symmetry is
imposed. Once we introduce a small rotation effect on the brane,
the bulk geometry should have a ``self-tuning'' process in order
to preserve the $Z_2$ symmetry. Therefore, some nontrivial
components of $ E_{\mu\nu}$ should be turned on to the order of
perturbation. Hence, the first constraint we need to check for a
consistent expression of $E_{\mu\nu}$ is the equation
(\ref{EqCon}).

For the perfect fluid source, two nontrivial equations of the
energy-momentum conservation (\ref{EqEMC}), corresponding to the
$t$ and $\phi$ components, are
\begin{eqnarray}
\dot \rho + 3 (\rho + p) \frac{\dot a}{a} &=& 0, \label{EqC1} \\
\partial_t \left[ (\rho + p) (\Omega - \omega) a^5 \right] &=& 0.
\label{EqC2}
\end{eqnarray}
Imposing the above equations, the right hand side of (\ref{EqCon})
has only one nonzero component as
\begin{equation}
D_\mu S^{\mu\phi} = - \frac16 \partial_t \left[ \rho (\rho + p) (\Omega -
\omega) a^5 \right] a^{-5}.
\end{equation}
Hence, the constraint (\ref{EqCon}) shows that the $E_{\mu\nu}$
should satisfy the following equations
\begin{equation}  \label{EqE}
D_\mu E^{\mu t} = D_\mu E^{\mu r} = D_\mu E^{\mu\theta} = 0, \qquad D_\mu
E^{\mu\phi} = - \frac{k_5^4}6 \partial_t \left[ \rho (\rho + p) (\Omega -
\omega) a^5 \right] a^{-5}.
\end{equation}
The last equation simply just shows what we have already
mentioned, that is the rotational perturbation does induce an
effect on the bulk for consistency. Otherwise, the only possible
perturbation is the so-called perfect dragging case, with
$\Omega=\omega$.

A possible solution of the constraint (\ref{EqE}) is $E^{\mu\nu}
= -(k_5^4/k_4^4) {\cal P}^{\mu\nu}$, which, for the perfect fluid
energy-momentum tensor gives only one non-vanishing component
\begin{equation}
{\cal P}^{t\phi} = {\cal P}^{\phi t} = {\cal P} \sim O(\epsilon),
\qquad D_\mu E^{\mu\phi} = - \frac{k_5^4}{k_4^4}
\partial_t \left( {\cal P} a^5 \right) a^{-5}.
\end{equation}
Hence the tensor component ${\cal P}$ is of order of perturbation
such that the quadratic terms, by itself or together with
$\omega$ or $\Omega$, can be neglected. Under this assumption,
the solution for the equation (\ref{EqE}) can be easily obtained
\begin{equation}
{\cal P}=\frac{k_4^4}6 \rho (\rho+p)(\Omega-\omega),
\end{equation}
which, as expected, is of the first order of perturbation.
Straightforwardly the first order of $E$-term is
\begin{equation}
E_{t\phi} = E_{\phi t} = \frac{k_5^4}6 a^2 r^2 \sin^2\theta \,
\rho (\rho + p) (\Omega - \omega),
\end{equation}
which is the consistently ``induced'' effect by the rotational
perturbation in the brane world model.

Generally, the above result can be extended to the case including
the scalar part of the $E$-correction, namely the ${\cal U}$ term
contribution. For the case, the first order covariant derivatives
are
\begin{eqnarray}
D_\mu E^{\mu t} &=& - \frac{k_5^4}{k_4^4} \left( \partial_t {\cal
U} + 4 \frac{\dot a}{a} {\cal U} \right), \qquad D_\mu E^{\mu r}
= D_\mu E^{\mu\theta} = 0, \nonumber \\
D_\mu E^{\mu\phi} &=& - \frac{k_5^4}{k_4^4} \partial_t \left[
-\frac43 {\cal U} (\Omega - \omega) a^5 + {\cal P} a^5 \right]
a^{-5} + \Omega D_\mu E^{\mu t}.
\end{eqnarray}
Then the constraint (\ref{EqE}) gives
\begin{equation}
{\cal U} = {\cal U}_0 a^{-4}, \qquad {\cal P} = \left[ \frac43
{\cal U} + \frac{k_4^4}6 \rho (\rho + p) \right] (\Omega -
\omega),
\end{equation}
or
\begin{equation}
E_{tt} = - \frac{k_5^4}{k_4^4} {\cal U}, \qquad E_{ii} = - \frac13
\frac{k_5^4}{k_4^4} {\cal U} g_{ii}, \qquad E_{t\phi} = E_{\phi
t} = a^2 r^2 \sin^2 \theta \left[ \frac13 \frac{k_5^4}{k_4^4}
{\cal U} \, \Omega + \frac{k_5^4}6 \rho (\rho+p)(\Omega-\omega)
\right].
\end{equation}

We assume that rotation is sufficiently slow, so that deviations
from spherical symmetry can be neglected. Then to first order in
$\Omega$ the gravitational and field equations (\ref{EqEinstein})
can be decomposed, by neglecting the ${\cal P}_{\mu\nu}$
contribution on the background geometry, into the following
components
\begin{eqnarray}
3 \frac{\dot a^2}{a^2} + \frac{3k}{a^2} &=& \Lambda + k_4^2 \rho
+ \frac{k_5^4}{12} \rho^2 + \frac{k_5^4}{k_4^4} {\cal U},
\label{Eqtt} \\
\frac{2\ddot a}{a} + \frac{\dot a^2}{a^2} + \frac{k}{a^2} &=&
\Lambda - k_4^2 p - \frac{k_5^4}{12} \rho (\rho + 2p) - \frac13
\frac{k_5^4}{k_4^4} {\cal U}, \label{Eqii}
\end{eqnarray}
\begin{equation}
3\frac{\dot a}{a} \frac{\partial \Omega}{\partial r} +
\frac{\partial^2 \Omega}{\partial t \partial r} = 0, \qquad
3\frac{\dot a}{a} \frac{\partial \Omega}{\partial \theta} +
\frac{\partial^2 \Omega}{\partial t \partial \theta} = 0,
\label{Eqrphi}
\end{equation}
\begin{eqnarray}
&& (1-kr^2) \frac{\partial^2 \Omega}{\partial r^2} + \left(
\frac4{r} - 5 k r \right) \frac{\partial \Omega}{\partial r} +
\frac1{r^2} \frac{ \partial^2 \Omega}{\partial \theta^2} + \frac{3
\cot\theta}{r^2} \frac{\partial \Omega }{\partial \theta} + 2 a^2
\Omega \left( \frac{2\ddot a}{a} + \frac{\dot a^2}{a^2} +
\frac{k}{a^2} \right) \nonumber
\\
&=& 2 a^2 \Omega \Lambda + 2 k_4^2 a^2 [(\Omega-\omega)\rho -
\omega p] + \frac{k_5^4}6 a^2 (\rho^2 \Omega - 2 \rho p \omega - 2
\rho^2 \omega) - 2 a^2 \left[ \frac13 \frac{k_5^4}{k_4^4} {\cal U}
\, \Omega + \frac{k_5^4}6 \rho (\rho+p)(\Omega-\omega) \right].
\label{Eqtphi1}
\end{eqnarray}
Equations (\ref{Eqrphi}) follows from the $(r\phi)$ and
$(\theta\phi)$ components of the field equations and
(\ref{Eqtphi1}) is from the $(t\phi)$-component. Moreover, the
last equation can be simplified by applying (\ref{Eqii}) to be
\begin{equation}
(1-kr^2) \frac{\partial^2 \Omega}{\partial r^2} + \left( \frac4{r}
- 5 k r \right) \frac{\partial \Omega}{\partial r} + \frac1{r^2}
\frac{ \partial^2 \Omega}{\partial \theta^2} + \frac{3
\cot\theta}{r^2} \frac{\partial \Omega }{\partial \theta} - 2
k_4^2 a^2 (\rho+p)(\Omega-\omega) = 0.  \label{Eqtphi}
\end{equation}
In this equation the correction from the bulk effect exactly
cancelled. Hence the form of this equation is the same as the
version from the standard general relativity.

The field equations (\ref{Eqtt})-(\ref{Eqtphi}) must be solved
together with the conservation equations
(\ref{EqC1})-(\ref{EqC2}). If the equation of state of the
cosmological matter, $p=p(\rho)$, is known, then Eq. (\ref{EqC1})
gives the time evolution of the energy density, $\rho=\rho(a)$.
Once $\rho$ is known, the general solution of the background
(unperturbed) gravitational field equations on the brane can be
obtained in the form
\begin{equation}
t-t_0 = \int \left[ \frac{\Lambda}3 a^2 + \frac{k_4^2}3 \rho(a)
a^2 + \frac{k_5^4}{36} \rho^2(a) a^2 + \frac{k_5^4}{3k_4^4}
\frac{{\cal U}_0}{a^2} - k \right]^{-1/2} da,
\end{equation}
where $t_0$ is a constant of integration.

The solution of equations (\ref{Eqrphi}) is
\begin{equation}
\Omega (t,r,\theta) = A(r,\theta) a^{-3}(t), \label{Omega}
\end{equation}
where $A(r,\theta)$ is an arbitrary integration function. Then the
perturbed equations we need to solve are
\begin{equation}
(1-kr^2) \frac{\partial^2 A}{\partial r^2} + \left( \frac4{r} - 5
k r \right) \frac{\partial A}{\partial r} + \frac1{r^2}
\frac{\partial^2 A}{\partial \theta^2} + \frac{3 \cot\theta}{r^2}
\frac{\partial A}{\partial \theta} - 2 k_4^2 a^2 (\rho+p)
(A-\omega a^3) = 0, \label{eq}
\end{equation}
and
\begin{equation}
\partial_t \left[ (\rho+p) (A-\omega a^3) a^2 \right] = 0.  \label{www}
\end{equation}

The general solution of Eq. (\ref{www}) is
\begin{equation}
A - \omega a^3 = \frac{F(r,\theta)}{(\rho+p) a^2},  \label{a}
\end{equation}
where $F(r,\theta)$ is an arbitrary integration function. With
the use of Eq. (\ref{Omega}) we obtain the following general
relation between the metric perturbation function $\Omega$ and
the angular velocity of the matter on the brane:
\begin{equation}
\omega (t,r,\theta) = \Omega(t,r,\theta) -
\frac{F(r,\theta)}{(\rho+p) a^5} = \frac{A(r,\theta)}{a^3} -
\frac{F(r,\theta)}{(\rho+p) a^5}. \label{omega}
\end{equation}

Finally, with the use of Eq. (\ref{a}) and of the unperturbed
field equations, Eq. (\ref{eq}) reduces to
\begin{equation}
(1-kr^2) \frac{\partial^2 A(r,\theta)}{\partial r^2} + \left(
\frac4{r} - 5 k r \right) \frac{\partial A(r,\theta)}{\partial r}
+ \frac1{r^2} \frac{\partial^2 A(r,\theta)}{\partial \theta^2} +
\frac{3 \cot\theta }{r^2} \frac{\partial A(r,\theta)}{\partial
\theta} - 2 k_4^2 F(r,\theta) = 0. \label{Eqfinal}
\end{equation}

\section{The case of the perfect dragging}
As a first case for the study of the rotational perturbations in
the brane world we consider the case of the perfect dragging,
corresponding to the choice $F(r,\theta)=0$ of the arbitrary
integration function. Therefore the angular velocity of the
matter equals the metric perturbation function and is given by
\begin{equation}
\omega(t,r,\theta) = A(r,\theta) a^{-3}(t).
\end{equation}
For an expanding Universe in the large time limit the rotational
perturbations always decay, $\omega$ tending to zero for $t \to
\infty$.

In order to find the radial and angular dependence of the angular
velocity, we consider two possible functional forms of the
function $A(r,\theta)$. As a first case we take $A(r,\theta)=B(r)
C(\theta)$. Then Eq. (\ref{Eqfinal}) can be separated into the
following two independent equations
\begin{eqnarray}
\frac{d^2 C}{d \theta^2} + 3 \cot\theta \frac{d C}{d \theta} + n
C &=& 0, \label{2} \\
(1-kr^2) r^2 \frac{d^2 B}{d r^2} + \left( \frac4{r} - 5 k r
\right) r^2 \frac{d B}{d r} - n B &=& 0,  \label{3}
\end{eqnarray}
where $n$ is a separation constant. By introducing a new variable
$x = \cos\theta$, Eq. (\ref{2}) becomes
\begin{equation}
(1 - x^2) \frac{d^2 C}{d x^2} - 4 x \frac{d C}{d x} + n C = 0.
\end{equation}

Therefore the general solution of Eq. (\ref{2}) is given by
\begin{equation}
C(\theta) = C_1 \, {}_2F_1(a_1,b_1;c_1;\cos^2\theta) + C_2
\cos\theta \, {}_2F_1(a_2,b_2;c_2;\cos^2\theta),
\end{equation}
where $a_1=(3-\sqrt{9+4n})/4, \, b_1=(3+\sqrt{9+4n})/4, \,
c_1=1/2, \, a_2=(5-\sqrt{9+4n})/4, \, b_2=(5+\sqrt{9+4n})/4, \,
c_2=3/2$ and ${}_2F_1(a,b;c;x)=\sum_{k=0}^{\infty} \frac{(a)_k
(b)_k}{(c)_k} \frac{x^k}{k!}$ is the hypergeometric function
\cite{AbSt72}. Hereafter in this section the parameters $C_1$ and
$C_2$ are used to label the constants of integration in the
expressions of exact solution.

In order to solve Eq. (\ref{3}) we introduce first a new variable
$\eta=r^2$. Hence the equation is transformed into
\begin{equation}
4 \eta^2 (1 - k \eta) \frac{d^2 B}{d \eta^2} + 2 \eta (5 - 6 k
\eta) \frac{d B}{d \eta} - n B = 0,
\end{equation}
with the general solution given by
\begin{equation}
B(r) = C_1 r^\alpha \, {}_2F_1(k_1,l_1;m_1;kr^2) + C_2 r^\beta \,
{}_2F_1(k_2,l_2;m_2;kr^2), \qquad \text{for} \quad k = \pm 1,
\end{equation}
where $\alpha=(-3-\sqrt{9+4n})/2, \, \beta=(-3+\sqrt{9+4n})/2, \,
k_1=-3/4-\sqrt{9+4n}/4, \, l_1=5/4-\sqrt{9+4n}/4, \,
m_1=1-\sqrt{9+4n}/2, \, k_2=-3/4+\sqrt{9+4n}/4, \,
l_2=5/4+\sqrt{9+4n}/4, \, m_2=1+\sqrt{9+4n}/2$. For $k=0$, the
general solution of Eq. (\ref{3}) is given by
\begin{equation}
B(r) = C_1 r^\alpha + C_2 r^\beta, \qquad \text{for} \quad k = 0.
\end{equation}

A second class of solutions can be obtained by assuming that the
function $A(r,\theta)$ can be represented as $A(r,\theta) = M(r)
+ N(\theta)$. Then Eq. (\ref{Eqfinal}) yields again two
independent equations,
\begin{eqnarray}
\frac{d^2 N}{d \theta^2} + 3 \cot\theta \frac{d N}{d \theta} &=&
- m, \label{7} \\
(1-kr^2) r^2 \frac{d^2 M}{d r^2} + \left( \frac4{r} - 5 k r
\right) r^2 \frac{d M}{d r} &=& m,  \label{6}
\end{eqnarray}
where $m$ is a separation constant. By introducing a new
variables $N'=d N/d \theta$, Eq. (\ref{7}) is transformed into
\begin{equation}
\frac{d N'}{d \theta} + 3 \cot\theta N' = - m.  \label{9}
\end{equation}

Taking integration, Eq. (\ref{9}) gives
\begin{equation}
N' = \frac{d N}{d \theta} = m \left(
\frac{\cos\theta}{\sin^3\theta} - \frac13 \cot^3\theta \right) +
\frac{C_1}{\sin^3\theta}.
\end{equation}
After further integration, we obtain the general solution of Eq.
(\ref{7}) in the form
\begin{equation}
N(\theta) = \frac{m}3 \left( \ln \sin\theta - \frac1{\sin^2\theta}
\right) + \frac{C_1}4 \left( \ln \frac{1-\cos\theta}{1+\cos\theta}
- \frac{2\cos\theta}{\sin^2\theta} \right) + C_2.
\end{equation}

The general solution of Eq. (\ref{6}), corresponding to the three
different background geometries is given by
\begin{equation}
M(r) = \left\{ \begin{array}{ll}
 - \frac1{6 r^3} \left[ m r + \sqrt{1+r^2} (2 r^2 - 1) (m
\sinh^{-1} r + C_1) - 2 r^3 (m \ln r + C_2) \right], \qquad &
\text{for} \quad k = -1, \\
 - \frac1{6r^3} \left[ - m r + \sqrt{1-r^2} (2 r^2 + 1) (m
\sin^{-1} r + C_1) - 2 r^3 (m \ln r + C_2) \right], & \text{for}
\quad k = 1, \\
 C_1 + \frac{C_2}{r^3} + \frac{m}3 \ln r, & \text{for} \quad k=0.
 \end{array} \right.
\end{equation}
Hence the general solution of the gravitational field equations
for a slowly rotating brane with perfect dragging can be obtained
in a closed form.

\section{General time dependence of the angular velocity for $\omega
\neq \Omega$}
Generally, the angular velocity of the matter on the slowly
rotating brane consists of two time-dependent terms,
$\omega(t,r,\theta)=\omega_1(t,r,\theta)+\omega_2(t,r,\theta)$.
The first term, $\omega_1(t,r,\theta)=A(r,\theta) a^{-3}(t)$
always tends to zero for an expanding Universe, with an
increasing scale factor, so that in the large time limit
$\omega_1 \to 0$. However, the time evolution of the second term,
$\omega_2(t,r,\theta)=-F(r,\theta)/(\rho+p) a^{5}$ essentially
depends on the equation of state and the dynamic behavior of the
background cosmological fluid and geometry. In the following we
shall investigate the time dependence of this term for a number
types of equation of state which could be relevant for the
description of the ultra-high density matter of which the
Universe consisted in its very early stages.

\subsection{Zeldovich Stiff Fluid}
One of the most common equations of state, which have extensively
been used to study the properties of the early Universe is the
linear barotropic equation of state, with $p=(\gamma-1)\rho$,
with $\gamma=\text{constant} \in [1,2]$. For this equation of
state the conservation equation (\ref{EqC1}) can be immediately
integrated to give $\rho=\rho_0 a^{-3\gamma}, \, \rho_0 =
\text{constant} \geq 0$. Therefore for the second term
$\omega_2(t,r,\theta)$ of the angular velocity we obtain the
following time dependence:
\begin{equation}
\omega_2(t,r,\theta) = - \frac{F(r,\theta)}{\gamma \rho_0} a^{3
\gamma - 5}.
\end{equation}

In order to have decaying rotational perturbations the condition
$\gamma<5/3$ must be satisfied. It is satisfied for the case of
the pressureless dust, with $\gamma=1$, and also for the
radiation fluid, having $\gamma=4/3$. But this condition in not
satisfied for a very important case of the so-called causal limit
of the linear barotropic equation of state, corresponding to
$\gamma=2$, or the Zeldovich stiff fluid equation of state
$p=\rho$. For the choice $\gamma=2$ we obtain
$\omega_2(t,r,\theta)=-F(r,\theta) a / 2 \rho_0$, showing that an
initial rotational perturbation in the ultra-high density
cosmological fluid do not decay and the angular velocity of the
matter is linearly increasing with the scale factor of the
expanding Universe.

The Zeldovich equation of state, valid for densities
significantly higher than nuclear densities, $\rho > 10
\rho_{\text{nuc}}$, with $\rho_{\text{nuc}}=10^{14} g/cm^3$ can
be obtained by constructing a relativistic Lagrangian that allows
bare nucleons to interact attractively via scalar meson exchange
and repulsively via the exchange of a more massive vector meson
\cite{ShTe83}. In the non-relativistic limit both the quantum and
classical theories yield Yukawa-type potentials. At the highest
densities the vector meson exchange dominates and by using a mean
field approximation one can show that in the extreme limit of
infinite densities the pressure tends to the energy density,
$p\to\rho$. In this limit the sound speed $c_s=\sqrt{dp/d\rho} \to
1$, and hence this equation of state satisfies the causality
condition, with the speed of sound less than the speed of light
\cite{ShTe83}.

The Zeldovich equation of state can also describe a
non-interacting scalar field, with zero potential, with
$\rho_\phi=p_\phi=\dot \phi^2/2$. Hence the rotational
perturbations in a potential-free field do not decay. For a
self-interacting field with potential $U(\phi)$, the energy
density and pressure of the field are $\rho_\phi=\dot \phi^2/2 +
U(\phi)$ and $p_\phi=\dot \phi^2/2 - U(\phi)$, respectively. The
conservation equation (\ref{EqC1}) becomes in this case
\begin{equation}
\ddot \phi + 3 \dot \phi \frac{\dot a}{a} + U'(\phi) = 0,
\end{equation}
and has the first integral
\begin{equation}
\dot \phi = \frac{K - \int a^3 U'(\phi) d t}{a^3},
\end{equation}
with $K$ a constant of integration. Therefore we obtain for
$\omega_2$
\begin{equation}
\omega_2 = - F(r,\theta) \frac{a}{K^2 - 2 K \int a^3 U'(\phi) d t
+ \left[ \int a^3 U'(\phi) d t \right]^2}.
\end{equation}
Hence if the potential $U(\phi)$ is so that the quantity $a \{
K^2 - 2 K \int a^3 U'(\phi) d t + [ \int a^3 U'(\phi) d t]^2
\}^{-1}$ is a decreasing function of time, the initial rotational
perturbation is rapidly decaying.

\subsection{Hagedorn Fluid}
An alternative approach to the equation of state at ultra-high
densities is based on the assumption that a whole host of baryonic
resonant states arise at high density \cite{ShTe83}. In the
Hagedorn model the baryon resonance mass spectrum is given by
$N(m) d m \sim m^a \exp(m/m_0) d m$, where $N(m) d m$ is the
number of resonances between mass $m$ and $m + d m$. The existing
data on baryon resonances show that $m_0=160$ MeV and $-7/2 \leq a
\leq -5/2$ \cite{LeWa73}. For asymptotically large densities the
particle number $n=\int_0^{\mu_n} N(m) d m \sim m_0 \mu_n^a \exp(
\mu_n/m_0)$, where $\mu_n$ is the chemical potential of the
nuclear matter. The density of the matter is $\rho \sim n \mu_n
\sim m_0 \mu_n^{a+1} \exp(\mu_n/m_0)$. The pressure can be
obtained from $p=n^2 d(\rho/n)/dn$ and is given by the Hagedorn
equation of state,
\begin{equation}
p = \frac{\rho}{\ln \frac{\rho}{\rho_0}}, \label{Hag}
\end{equation}
where $\rho_0=2.5\times 10^{12}$ g/cm$^3$ \cite{ShTe83}.

The velocity of sound in this type of matter is $c_s=[
\ln(\rho/\rho_0)]^{-1/2} \left[ 1 - ( \ln(\rho/\rho_0) )^{-1}
\right]^{1/2}$ \cite{LeWa73,ShTe83}. For the Hagedorn equation of
state the speed of sound has the property $c_s \to 0$ for
$\rho/\rho_0 \to \infty$, in striking contrast with the mean field
theory approach in which $c_s \to 1$. The gravitational collapse
of a high-density null charged matter fluid, satisfying the
Hagedorn equation of state in the framework of the Vaidya geometry
was considered in \cite{Ha03}. A collapsing Hagedorn fluid could
end either as a black hole, or as a naked singularity. The
collapse of Hagedorn fluid to a naked singularity is also a
possible source of gamma-ray bursts \cite{Ha03}.

In order to find the time behavior of the rotational
perturbations in a Hagedorn fluid, with the equation of state
given by (\ref{Hag}), we have to obtain first the scale factor
dependence of the density. For the Hagedorn equation of state the
conservation equation of the matter on the brane (\ref{EqC1})
takes the form
\begin{equation}
\dot \rho + 3 \rho \left[ 1 + \left( \ln \frac{\rho}{\rho_0}
\right)^{-1} \right] \frac{\dot a}{a} = 0,
\end{equation}
which can be integrated to give
\begin{equation}
\frac{\rho}{\rho_0 \left( \ln \frac{\rho}{\rho_0} + 1 \right)} =
\frac{C}{a^3},
\end{equation}
with $C\geq 0$ a constant of integration. For an expanding
Universe with an increasing scale factor the energy density of
the Hagedorn cosmological fluid is decreasing in time. The time
variation of the second component of the angular velocity of the
matter can be written as a function of the time dependent only
density as
\begin{equation}
\omega_2(t,r,\theta) = - \frac{F(r,\theta)}{\rho_0 C^{5/3}}
\frac{\left( \frac{\rho}{\rho_0} \right)^{2/3} \ln
\frac{\rho}{\rho_0}}{\left( \ln \frac{\rho}{\rho_0} + 1
\right)^{8/3}}. \label{omhag}
\end{equation}

The variation of the time-dependent part $f(\rho/\rho_0) =
(\rho/\rho_0)^{2/3} \ln (\rho/\rho_0) / \left[ \left( \ln
(\rho/\rho_0) + 1 \right)^{8/3} \right]$ of $\omega_2$ is
represented in Fig. 1.

\vspace{0.2in}
\begin{figure}[h]
\includegraphics{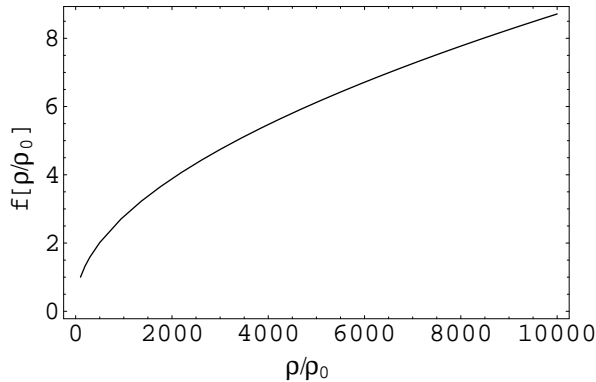}
\caption{Variation, as a function of $\rho/\rho_0$, of the time
dependent part $f(\rho/\rho_0)$ of $\omega_2$.} \label{FIG1}
\end{figure}

As one can see from the figure, once the Universe is expanding
and its density decreases, the angular velocity of the matter is
also decreasing, and in the large time limit it tends to zero.
Therefore the cosmological behavior of the rotational
perturbations of the Hagedorn fluid is very different to that of
the causal stiff (Zeldovich) cosmological matter, with the matter
angular velocity increasing due to the cosmological expansion.

\subsection{Chaplygin Gas}
A form of matter which has recently been invoked to account for
the recent supernovae evaluations of the cosmological constant is
the so-called Chaplygin gas \cite{GiHoYi01}. The Chaplygin gas is
a form of what is called ``k-essence'', and satisfies an equation
of state of the form $p=-1/\rho$ \cite{Gi01}. This equation of
state comes from the Born-Infeld type Lagrangian, $\int (1 -
\sqrt{1 - \eta^{\mu\nu} \partial_\mu A \partial_\nu A}) d^n x$,
with $A$ the transverse coordinate of domain wall or a
(n-1)-brane. The energy-momentum tensor satisfies the Hooke law,
and, by taking into account a cosmological constant $\lambda_0$,
it follows that the pressure and the energy density satisfy the
equation of state $(\rho+\lambda_0)(p-\lambda_0)=-1$ \cite{Gi01}.
In the following we shall take $\lambda_0=0$. Hence the energy
conservation equation for a Chaplygin gas on the brane becomes
\begin{equation}
\dot \rho + 3 \left( \rho - \frac1{\rho} \right) \frac{\dot a}{a}
= 0,
\end{equation}
with the general solution given by
\begin{equation}
\rho = \sqrt{\frac{\rho_0}{a^6} + 1},
\end{equation}
with $\rho_0 \geq 0$ a constant of integration.

Therefore the second component of the angular velocity varies as
\begin{equation}
\omega_2(t,r,\theta) = - \frac{F(r,\theta)}{\rho_0}
\frac{\sqrt{a^6+\rho_0}}{a^2}.
\end{equation}

In the limit of large times $a^6>>\rho_0$ and it follows that
$\omega_2(t,r,\theta)\to -F(r,\theta) a/\rho_0$, that is, similar
to the Zeldovich fluid, the angular velocity of the expanding
Universe is an increasing function of time.

\subsection{Classical Boltzmann Gas}
Another important example of a rotationally perturbed
cosmological fluid we consider next is the classical Boltzmann
gas filled Universe. If the cosmological fluid is a
collision-dominated classical gas in equilibrium, then the
thermodynamic parameters of the gas are given by \cite{Hi91},
\cite{Ma96}
\begin{equation}
p = n m \beta^{-1}, \qquad n m = A_0 K_2(\beta) / \beta, \qquad
\rho = A_0 \left[ \beta^{-1} K_1(\beta) + 3 \beta^{-2} K_2(\beta)
\right],  \label{Boltz}
\end{equation}
where $n$ is the particle number density, $\beta=m c^2/k_B T$,
with $k_B$ the Boltzmann constant, $T$ the temperature and
$A_0=m^4 g / 2\pi^2 \hbar^3 $, with $g$ the spin weight of the
fluid particles. $K_n$ are the modified Bessel functions of the
second kind.

With the use of Eqs. (\ref{Boltz}) the energy conservation
equation of the cosmological fluid can be written as
\begin{equation}
\dot \beta - g(\beta) \frac{\dot a}{a} = 0,
\end{equation}
where
\begin{equation}
g(\beta) = \frac{6\beta [ \beta K_1(\beta) + 4 K_2(\beta) ]}
{\beta^2 K_0(\beta) + 5 \beta K_1(\beta) + ( 12 + \beta^2)
K_2(\beta) + 3 \beta K_3(\beta)}. \label{fung}
\end{equation}
In order to obtain Eq. (\ref{fung}) we have used the recurrence
relation $K'_n(x)=-[ K_{n-1}(x)+K_{n+1}(x)]/2$ \cite{AbSt72}.
Therefore the scale factor can be expressed as a function of the
temperature as
\begin{equation}
a(\beta) = a_0 \exp \left( \int \frac{d \beta}{g(\beta)}\right),
\end{equation}
with $a_0 \geq 0$ a constant of integration.

The second component of the angular velocity of the matter on the
brane is given by
\begin{equation}
\omega_2(\beta,r,\theta) = - \frac{F(r,\theta)} {A_0 a_0^5}
\frac{\beta^2 \exp \left( -5 \int g^{-1}(\beta) \, d \beta
\right)}{\beta K_1(\beta) + 4 K_2(\beta)} = - \frac{F(r,\theta)}
{A_0 a_0^5} \, h(\beta),
\end{equation}
where we denoted $h(\beta)=\beta^2\exp \left[ -5 \int
g^{-1}(\beta) \, d\beta \right] / [ \beta K_1(\beta) + 4
K_2(\beta) ]$. The variation of $h(\beta)$ is presented in Fig. 2.

\vspace{0.2in}
\begin{figure}[h]
\includegraphics{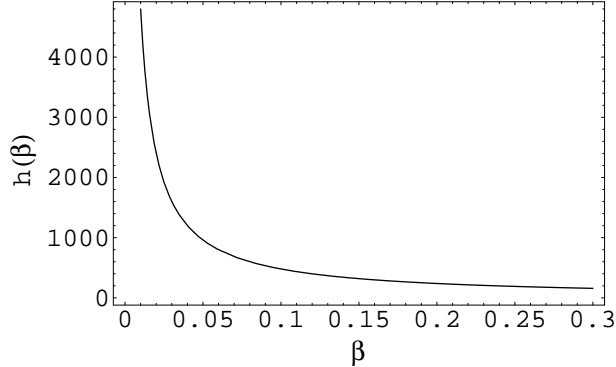}
\caption{Variation, as a function of $\beta$, of the time
dependent part $h(\beta)$ of $\omega_2$.} \label{FIG2}
\end{figure}

The angular velocity of the Boltzmann gas decreases with
increasing $\beta$, so that for small temperatures (large $\beta
$) $\omega_2\to 0$.

\subsection{Quark Matter}
The rotational perturbations always decay in the important case
of quark matter, satisfying the bag model equation of state
$p=(\rho-4B)/3$ \cite{Wi84}, where $B=10^{14} g/cm^3$ is the bag
constant \cite{ChHa00}. Quark matter is assumed to have played an
important role in the early evolution of the Universe, and a
first order phase transition from quark to hadronic matter took
place when the temperature was around $100 MeV$. It is also
possible that inhomogeneities in the baryon number density were
produced during this transition, and perhaps they persisted even
to the time of nucleosynthesis, which could alter the abundance
of light elements \cite{Wi84}. For the bag model equation of
state of the quark matter the density-scale factor dependence is
given by
\begin{equation}
\rho = B + \frac{\rho_0}{a^4},
\end{equation}
and consequently $\omega_2$ becomes
\begin{equation}
\omega_2(t,r,\theta) = - \frac{3 F(r,\theta)}{4\rho_0} \frac1{a}.
\end{equation}
Therefore in the large time limit, for $a \to \infty$, the
rotational perturbations in the cosmological quark matter tend to
zero.

In the general case, due to the dependence of the rotation
equation on the arbitrary function $F(r,\theta)$, the solution of
the rotation equation cannot be obtained. The particular choice
$F(r,\theta)=A(r,\theta)$ leads again to a radial and angular
distribution of the matter which is very similar to the perfect
dragging case. Moreover, if $A$ is represented as a product of
two $r$ and $\theta$ only dependent functions, $A(r,\theta) =
X(r) Y(\theta)$, then the angular velocity $\omega$ is also a
separable function of all variables. But if $A$ is the sum of two
independent functions, $A(r,\theta) = X(r) + Y(\theta)$, the
angular velocity is a non-separable function. In these cases the
functions $X(r)$ and $Y(\theta)$ can be obtained in terms of the
hypergeometric function.

\section{Discussions and final remarks}
In the present paper we have considered the evolution of the small
rotational perturbations in the brane world cosmological model.
The evolution of the angular velocity of the matter is related to
the metric perturbation function via an equation derived from the
energy-momentum conservation on the brane. This equation is in
fact the same in both standard general relativity and brane world
cosmology. Once the evolution of the background non-perturbed
geometry is known, the time dynamics of the metric rotation
function $\Omega(t,r,\theta)$ and of the angular velocity
$\omega(t,r,\theta)$ of the matter on the brane is uniquely
determined by the unperturbed field equations and, from a
physical point of view, by the equation of state of the
ultra-high density cosmological fluid. The time behaviors of
$\omega(t,r,\theta)$ and $\Omega(t,r,\theta)$ are independent on
their spatial distribution. For $\omega=\Omega$, in the large
time limit the angular velocity of the matter tends to zero,
showing that the initial small rotational perturbations of the
brane world are smoothed out, due to the expansion of the
Universe. This result is also independent of the geometry (flat,
open or close) of the space-time. However, despite the fact that
the presence of the dark matter term ${\cal U}={\cal U}_0/a^4$ is
not essential for the rapid decay of the rotational perturbations
of the matter on the brane, the effects induced by the
five-dimensional bulk still contribute to the decay (or
amplification) of the first order rotational perturbations via
their effect on the unperturbed background geometry.

A very different situation occurs for $\omega \neq \Omega$. In
this case the dependence of the angular velocity on the equation
of state leads to two distinct classes of behaviors. For some
equations of state (radiation fluid or quark matter), the
rotational perturbations decay and this behavior is consistent
with the observational constraint that our present day Universe is
rotating very slowly. However, some equations of state which try
to model the thermodynamical properties of the super-dense matter
at the very early stages of the evolution of our Universe, like
the Zeldovich or Chaplygin gas equations of state, do not lead to
an observationally consistent description of the early Universe.
In particular, the causal Zeldovich equation of state with
$\gamma=2$, which has been used, for example, to determine the
maximum mass of the neutron stars \cite{ShTe83}, does not lead,
once applied to a cosmological framework, to a description of the
rotational perturbations consistent with the observations, that
is, it does not satisfy the basic observational requirement of a
very small rotation in the large time limit. Hence, if still one
assumes these equations of state for the early Universe, a
physical mechanism supplementary to the expansion is also needed
to suppress rotation.

However, if the brane Universe experiences an inflationary period,
than any rotational perturbation is reduced exponentially to a
very low level, the decrease of cosmological vorticity being of
the order of $10^{-145}$ \cite{GrSo87}. Hence, any growth of the
rotational perturbations during a maximally stiff phase in a
post-inflationary period is unlikely to increase them back up to
observationally significant levels.

The equation of state of the Hagedorn fluid is consistent with the
requirement of a non-rotating Universe. Therefore the study of the
perturbations of cosmological models could also provide some
insights in the equation of state of the ultra-high density
nuclear or sub-nuclear matter. The actual very tight limit on the
rotation of the Universe imposes very strong constraints on the
initial equation of state of the cosmological fluid.

The spatial and angular distribution of the angular velocity
depend on two arbitrary integration constants $C_i, i=1,2$. In
principle the values of these constants could be determined by
fixing the value of $A(r,\theta)$ at the center, and from the
actual value of the angular velocity of the Universe.

\section*{Acknowledgments}
The authors would like to thank to the anonymous referee for
comments which helped to improve the manuscript. This work is
supported in part by the National Science Council under the grant
numbers NSC90-2112-M009-021 and NSC90-2112-M002-055. The work of
CMC is also supported by the Taiwan CosPA project and, in part, by
the Center of Theoretical Physics at NTU and National Center for
Theoretical Science.


\begin{references}

\bibitem{Bi82}
  P. Birch,
  {\sl Is the universe rotating?},
  {\it Nature \bf 298} (1982) 451-454.

\bibitem{Bi83}
  P. Birch,
  {\sl Is there evidence for universal rotation? Birch replies},
  {\it Nature \bf 301} (1982) 736.

\bibitem{NoRa97}
  B. Nodland and J. P. Ralston,
  {\sl Indication of anisotropy in electromagnetic propagation over
       cosmological distances},
  {\it Phys. Rev. Lett. \bf 78} (1997) 3043-3046; {\tt astro-ph/9704196}.

\bibitem{Ku97}
  R. W. K\"uhne,
  {\sl On the cosmic rotation axis},
  {\it Mod. Phys. Lett. \bf A12} (1997) 2473-2474; {\tt astro-ph/9708109}.

\bibitem{CoHa73}
  C. B. Collins and S. W. Hawking,
  {\sl The rotation and distortion of the universe},
  {\it Mon. Not. R. Astr. Soc. \bf 162} (1973) 307-320.

\bibitem{BaJuSo85}
  J. D. Barrow, R. Juszkiewicz and D. H. Sonoda,
  {\sl Universal rotation: how large can it be},
  {\it Mon. Not. R. Astr. Soc. \bf 213} (1985) 917-943.

\bibitem{Go49}
  K. G\"odel,
  {\sl An example of a new type of cosmological solutions of
       Einstein's field equations for gravitation},
  {\it Rev. Mod. Phys. \bf 21} (1949) 447-450.

\bibitem{Ob00}
  Yu. N. Obukhov,
  {\sl On physical foundations and observational effects of cosmic rotation},
  in: {\it Colloquium on Cosmic Rotation},
  eds. M. Scherfner, T. Chrobok and M. Shefaat,
  Wissenschaft und Technik Verlag, Berlin (2000) 23-96; {\tt astro-ph/0008106}.

\bibitem{ElOl83}
  J. Ellis and K. A. Olive,
  {\sl Inflation can solve the rotation problem},
  {\it Nature \bf 303} (1983) 679-681.

\bibitem{GrSo87}
  {\O }. Gr{\o }n and H. H. Soleng,
  {\sl Decay of primordial cosmic rotation in inflationary cosmologies},
  {\it Nature \bf 328} (1987) 501-503.

\bibitem{Ha66}
  S. W. Hawking,
  {\sl Perturbations of an expanding Universe},
  {\it Astrophys. J. \bf 145} (1966) 544-554.

\bibitem{Mi79}
  A. Miyazaki,
  {\sl Dragging effect on the inertial frame and the contribution
       of matter to the gravitational ``constant'' in a closed
       cosmological model of the Brans-Dicke theory},
  {\it Phys. Rev. \bf D19} (1979) 2861-2867.

\bibitem{BaCo80}
  S. S. Bayin and F. I. Cooperstock,
  {\sl Rotational perturbations of Friedmann Universes},
  {\it Phys. Rev. \bf D22} (1980) 2317-2322.

\bibitem{Ba85}
  S. S. Bayin,
  {\sl Comments on rotational perturbations of Friedmann models},
  {\it Phys. Rev. \bf D32} (1985) 2241-2242.

\bibitem{ChHaMa01}
  C.-M. Chen, T. Harko and M. K. Mak,
  {\sl Rotational perturbations in Neveu-Schwarz-Neveu-Schwarz
       string cosmology},
  {\it Phys. Rev. \bf D63} (2001) 104013; {\tt hep-th/0012151}.

\bibitem{RS99a}
  L. Randall and R. Sundrum,
  {\sl A large mass hierarchy from a small extra dimension},
  {\it Phys. Rev. Lett. \bf 83} (1999) 3370-3373; {\tt hep-ph/9905221}.

\bibitem{RS99b}
  L. Randall and R. Sundrum,
  {\sl An alternative to compactification},
  {\it Phys. Rev. Lett \bf 83} (1999) 4690-4693; {\tt hep-th/9906064}.

\bibitem{HW96}
  P. Horava and E. Witten,
  {\sl Heterotic and type I string dynamics from eleven-dimensions},
  {\it Nucl. Phys. \bf B460} (1996) 506-524; {\tt hep-th/9510209}.

\bibitem{KK00}
  H. B. Kim and H. D. Kim,
  {\sl Inflation and gauge hierarchy in Randall-Sundrum compactification},
  {\it Phys. Rev. \bf D61} (2000) 064003; {\tt hep-th/9909053}.

\bibitem{BDEL00}
  P. Binetruy, C. Deffayet, U. Ellwanger and D. Langlois,
  {\sl Brane cosmological evolution in a bulk with cosmological constant},
  {\it Phys. Lett. \bf B477} (2000) 285-291; {\tt hep-th/9910219}.

\bibitem{BDL00}
  P. Binetruy, C. Deffayet and D. Langlois,
  {\sl Nonconventional cosmology from a brane universe},
  {\it Nucl. Phys. \bf B565} (2000) 269-287; {\tt hep-th/9905012}.

\bibitem{STW00}
  H. Stoica, S.-H. H. Tye and I. Wasserman,
  {\sl Cosmology in the Randall-Sundrum brane world scenario},
  {\it Phys. Lett. \bf B482} (2000) 205-212; {\tt hep-th/0004126}.

\bibitem{LMW00}
  D. Langlois, R. Maartens and D. Wands,
  {\sl Gravitational waves from inflation on the brane},
  {\it Phys. Lett. \bf B489} (2000) 259-267; {\tt hep-th/0006007}.

\bibitem{CEHS00}
  C. Csaki, J. Erlich, T. J. Hollowood and Y. Shirman,
  {\sl Universal aspects of gravity localized on thick branes},
  {\it Nucl. Phys. \bf B581} (2000) 309-338; {\tt hep-th/0001033}.

\bibitem{ClGrSe99}
  J. M. Cline, C. Grojean and G. Servant,
  {\sl Cosmological expansion in the presence of extra dimensions},
  {\it Phys. Rev. Lett. \bf 83} (1999) 4245; {\tt hep-ph/9906523}.

\bibitem{AnNuOl00}
  L. Anchordoqui, C. Nunez and K. Olsen,
  {\sl Quantum cosmology and ADS/CFT},
  {\it JHEP \bf 0010} (2000) 050; {\tt hep-th/0007064}.

\bibitem{Ma01}
  R. Maartens,
  {\sl Geometry and dynamics of the brane-world},
  {\tt gr-qc/0101059}.

\bibitem{SMS00}
  T. Shiromizu, K. Maeda and M. Sasaki,
  {\sl The Einstein equation on the 3-brane world},
  {\it Phys. Rev. \bf D62} (2000) 024012; {\tt gr-qc/9910076}.

\bibitem{SSM00}
  M. Sasaki, T. Shiromizu and K. Maeda,
  {\sl Gravity, stability and energy conservation on the
       Randall-Sundrum brane world},
  {\it Phys. Rev. \bf D62} (2000) 024008; {\tt hep-th/9912233}.

\bibitem{MW00}
  K. Maeda and D. Wands,
  {\sl Dilaton gravity on the brane},
  {\it Phys. Rev. \bf D62} (2000) 124009; {\tt hep-th/0008188}.

\bibitem{MSS00}
  R. Maartens, V. Sahni and T. D. Saini,
  {\sl Anisotropy dissipation in brane world inflation},
  {\it Phys.Rev. \bf D63} (2001) 063509; {\tt gr-qc/0011105}.

\bibitem{CS01}
  A. Campos and C. F. Sopuerta,
  {\sl Evolution of cosmological models in the brane world scenario},
  {\it Phys. Rev. \bf D63} (2001) 104012; {\tt hep-th/0101060}.

\bibitem{CS01a}
  A. Campos and C. F. Sopuerta,
  {\sl Bulk effects in the cosmological dynamics of brane world scenarios},
  {\it Phys. Rev. \bf D64} (2001) 104011; {\tt hep-th/0105100}.

\bibitem{ChHaMa01a}
  C.-M. Chen, T. Harko and M. K. Mak,
  {\sl Exact anisotropic brane cosmologies},
  {\it Phys. Rev. \bf D64} (2001) 044013; {\tt hep-th/0103240}.

\bibitem{Co01a}
  A. Coley,
  {\sl Dynamics of brane-world cosmological models},
  {\it Phys. Rev. \bf D66} (2002) 023512;
  {\tt hep-th/0110049}.

\bibitem{Co01b}
  A. Coley,
  {\sl No chaos in brane-world cosmology},
    {\it Class. Quantum Grav. \bf 19} (2002) L45-L49; {\tt
    hep-th/0110117}.

\bibitem{ChHaMa01b}
  C.-M. Chen, T. Harko and M. K. Mak,
  {\sl Viscous dissipative effects in isotropic brane cosmology},
  {\it Phys. Rev. \bf D64} (2001) 124017; {\tt hep-th/0106263}.

\bibitem{BaMa01}
  J. D. Barrow and R. Maartens,
  {\sl Kaluza-Klein anisotropy in the CMB},
  {\it Phys. Lett. \bf B532} (2002) 153-158; {\tt gr-qc/0108073}.

\bibitem{BaHe01}
  J. D. Barrow and S. Hervik,
  {\sl Magnetic brane-worlds},
  {\it Class. Quantum Grav. \bf 19} (2002) 155-172; {\tt gr-qc/0109084}.

\bibitem{HaMa03}
T. Harko and M. K. Mak, {\sl Viscous Bianchi type I universes in
brane cosmology}, {\it Class. Quantum Grav. \bf 20} (2003)
407-422; {\tt gr-qc/0212075}.

\bibitem{Ma00}
  R. Maartens,
  {\sl Cosmological dynamics on the brane},
  {\it Phys. Rev. \bf D62} (2000) 084023; {\tt hep-th/0004166}.

\bibitem{KoIsSa00}
  H. Kodama, A. Ishibashi and O. Seto,
  {\sl Brane world cosmology: gauge invariant formalism for perturbation},
  {\it Phys. Rev. \bf D62} (2000) 064022; {\tt hep-th/0004160}.

\bibitem{Ko01}
  H. Kodama,
  {\sl Behavior of cosmological perturbations in the brane world
       model};
  {\tt hep-th/0012132}.

\bibitem{La00}
  D. Langlois,
  {\sl Brane cosmological perturbations},
  {\it Phys. Rev. \bf D62} (2000) 126012; {\tt hep-th/0005025}.

\bibitem{La01}
  D. Langlois,
  {\sl Evolution of cosmological perturbations in a
       brane-universe},
  {\it Phys. Rev. Lett. \bf 86} (2001) 2212-2215; {\tt hep-th/0010063}.

\bibitem{BrDoBrLu00}
  C. van de Bruck, M. Dorca, R. H. Brandenberger and A. Lukas,
  {\sl Cosmological perturbations in brane world theories: formalism},
  {\it Phys. Rev. \bf D62} (2000) 123515; {\tt hep-th/0005032}.

\bibitem{KoSo00}
  K. Koyama and J. Soda,
  {\sl Evolution of cosmological perturbations in the brane world},
  {\it Phys. Rev. \bf D62} (2000) 123502; {\tt hep-th/0005239}.

\bibitem{GoMa00}
  C. Gordon and R. Maartens,
  {\sl Density perturbations in the brane world},
  {\it Phys. Rev. \bf D63} (2001) 044022; {\tt hep-th/0009010}.

\bibitem{LaMaSaWa01}
  D. Langlois, R. Maartens, M. Sasaki and D. Wands,
  {\sl Large-scale cosmological perturbations on the brane},
  {\it Phys. Rev. \bf D63} (2001) 084009; {\tt hep-th/0012044}.

\bibitem{DeDoKa01}
  N. Deruelle, T. Dolezel and J. Katz,
  {\sl Perturbations of brane worlds},
  {\it Phys. Rev. \bf D63} (2001) 083513; {\tt hep-th/0010215}.

\bibitem{BrDo01}
  M. Dorca and C. van de Bruck,
  {\sl Cosmological Perturbations in Brane Worlds: Brane Bending and Anisotropic
  Stresses},
  {\it Nucl. Phys. \bf B605} (2001) 215-233;
  {\tt hep-th/0012073}.

\bibitem{KuTa01}
  H. Kudoh and T. Tanaka,
  {\sl Second order perturbations in the Randall-Sundrum infinite
       brane world model},
  {\it Phys. Rev. \bf D64} (2001) 084022; {\tt hep-th/0104049}.

\bibitem{KuTa01a}
  H. Kudoh and T. Tanaka,
  {\sl Second order perturbations in the radius stabilized Randall-Sundrum
       two branes model},
  {\it Phys. Rev. \bf D65} (2002) 104034; {\tt hep-th/0112013}.

\bibitem{BrMaWa01}
  H. A. Bridgman, K. A. Malik and D. Wands,
  {\sl Cosmological perturbations in the bulk and on the brane},
  {\it Phys. Rev. \bf D65} (2002) 043502; {\tt astro-ph/0107245}.

\bibitem{BrBrDa01}
  P. Brax, C. van de Bruck and A. C. Davis,
  {\sl Brane world cosmology, bulk scalars and perturbations},
  {\it JHEP \bf 0110} (2001) 026; {\tt hep-th/0108215}.

\bibitem{Ma03} R. Maartens,
{\sl Brane-world cosmological perturbations: a covariant
approach}, {\it Prog. Theor. Phys. Suppl. \bf 148} (2002) 213;
{\tt gr-qc/0304089}.

\bibitem{ChHaKaMa02}
  C.-M. Chen, T. Harko, W. F. Kao and M. K. Mak,
  {\sl Rotational perturbations of brane world cosmological models},
  {\it Nucl. Phys. \bf B64} (2002) 159-178; {\tt hep-th/0201012}.

\bibitem{MuFeBr92}
  V. F. Mukhanov, F. A. Feldman and R. H. Brandenberger,
  {\sl Theory of cosmological perturbations. Part 1. Classical
       perturbations. Part 2. Quantum theory of perturbations. Part 3.
       Extensions},
  {\it Phys. Rep. \bf 215} (1992) 203-333.

\bibitem{AbSt72}
  M. Abramowitz and I. A. Stegun,
  {\sl Handbook of Mathematical Functions},
  Washington D. C., National Bureau of Standards, (1972).

\bibitem{ShTe83}
  S. L. Shapiro and S. A. Teukolsky,
  {\sl Black holes, white dwarfs and neutron stars},
  New York, John Wiley \& Sons, (1983).

\bibitem{LeWa73}
  Y. C. Leung and C. G. Wang,
  {\sl Equation of state of matter at supernuclear densities},
  {\it Astrophys. J. \bf 181} (1973) 895-902.

\bibitem{Ha03}
T. Harko, {\sl Gravitational collapse of a Hagedorn fluid in
Vaidya geometry},  {\it Phys. Rev. \bf D68} (2003) 064005; {\tt
gr-qc/0307064}.

\bibitem{GiHoYi01}
  G. W. Gibbons, K. Hori and P. Yi,
  {\sl String fluid from unstable D-branes},
  {\it Nucl. Phys. \bf B596} (2001) 136-150; {\tt hep-th/0009061}.

\bibitem{Gi01}
  G. W. Gibbons,
  {\sl Pulse propagation in Born-Infeld theory, the world-volume
       equivalence principle and the Hagedorn like equation of state of
       the Chaplygin gas}, {\sl Grav. Cosmol. \bf 8} (2002) 2-6;
    {\tt hep-th/0104015}.

\bibitem{Hi91}
  W. A. Hiscock and J. Salmonson,
  {\sl Dissipative Boltzmann-Robertson-Walker cosmologies},
  {\it Phys. Rev. \bf D43} (1991) 3249-3258.

\bibitem{Ma96}
  R. Maartens,
  {\sl Causal thermodynamics in relativity};
  {\tt astro-ph/9609119}.

\bibitem{Wi84}
  E. Witten,
  {\sl Cosmic separation of phases},
  {\it Phys. Rev. \bf D30} (1984) 272-285.

\bibitem{ChHa00}
  K. S. Cheng and T. Harko,
  {\sl Approximate mass and radius formulas for static and rotating
       strange stars},
  {\it Phys. Rev. \bf D62} (2000) 083001.

\end{references}
\end{document}